# Stability of vortex solitons in thermal nonlinear media with cylindrical symmetry

Yaroslav V. Kartashov,[1] Victor A. Vysloukh,[2] and Lluis Torner[1]

[1]*ICFO-Institut de Ciencies Fotoniques, and Universitat Politecnica de Catalunya, Mediterranean Technology Park, 08860 Castelldefels* (*Barcelona*), *Spain*
[2]*Departamento de Fisica y Matematicas, Universidad de las Americas – Puebla, Santa Catarina Martir, 72820, Puebla, Mexico*
*Yaroslav.Kartashov@icfo.es*

**Abstract:** We analyze the salient features of vortex-ring solitons supported by cylindrically symmetric media with nonlocal thermal nonlinearity. We discover the existence of a maximum allowed topological charge for such vortex solitons to be stable on propagation: Only vortex-ring solitons with topological charge $m \leq 2$ are found to be stable. This remarkable result holds independently of the radius of the sample.

**References and links**

1. W. Krolikowski, O. Bang, N. I. Nikolov, D. Neshev, J. Wyller, J. J. Rasmussen, and D. Edmundson, "Modulational instability, solitons and beam propagation in spatially nonlocal nonlinear media," J. Opt. B: Quantum and Semiclass. Opt. **6**, S288 (2004).
2. C. Conti, M. Peccianti, and G. Assanto, "Route to nonlocality and observation of accessible solitons," Phys. Rev. Lett. **91**, 073901 (2003).
3. C. Conti, M. Peccianti, and G. Assanto, "Observation of optical spatial solitons in highly nonlocal medium," Phys. Rev. Lett. **92**, 113902 (2004).
4. M. Peccianti, C. Conti, and G. Assanto, "Interplay between nonlocality and nonlinearity in nematic liquid crystals," Opt. Lett. **30**, 415 (2005).
5. V. A. Mironov, A. M. Sergeev, and E. M. Sher, "Non-unidimensional coupled solitons in field non-linear equations," Dokl. Akad. Nauk SSSR **260**, 325 (1981).
6. S. Lopez-Aguayo, A. S. Desyatnikov, Y. S. Kivshar, S. Skupin, W. Krolikowski, and O. Bang, "Stable rotating dipole solitons in nonlocal optical media," Opt. Lett. **31**, 1100 (2006).
7. Y. V. Kartashov, L. Torner, V. A. Vysloukh, and D. Mihalache, "Multipole vector solitons in nonlocal nonlinear media," Opt. Lett. **31**, 1483 (2006).
8. C. Rotschild, M. Segev, Z. Xu, Y. V. Kartashov, L. Torner, and O. Cohen, "Two-dimensional multipole solitons in nonlocal nonlinear media," Opt. Lett. **31**, 3312 (2006).
9. S. Lopez-Aguayo, A. S. Desyatnikov, and Y. S. Kivshar, "Azimuthons in nonlocal nonlinear media," Opt. Express **14**, 7903 (2006).
10. V. I. Kruglov, Y. A. Logvin, and V. M. Volkov, "The theory of spiral laser beams in nonlinear media," J. Mod. Opt. **39**, 2277 (1992).
11. D. Briedis, D. Petersen, D. Edmundson, W. Krolikowski, and O. Bang, "Ring vortex solitons in nonlocal nonlinear media," Opt. Express **13**, 435 (2005).
12. A. I. Yakimenko, Y. A. Zaliznyak, and Y. Kivshar, "Stable vortex solitons in nonlocal self-focusing nonlinear media," Phys. Rev. E **71**, 065603(R).
13. A. M. Deykoon and G. A. Swartzlander, "Pinched optical-vortex soliton," J. Opt. Soc. Am. B **18**, 804 (2001).
14. S. Skupin, O. Bang, D. Edmundson, and W. Krolikowski, "Stability of two-dimensional spatial solitons in nonlocal nonlinear media," Phys. Rev. E **73**, 066603 (2006).
15. A. I. Yakimenko, V. M. Lashkin, and O. O. Prikhodko, "Dynamics of two-dimensional coherent structures in nonlocal nonlinear media," Phys. Rev. E **73**, 066605 (2006).
16. A. S. Desyatnikov, Y. S. Kivshar, L. Torner, "Optical vortices and vortex solitons," Prog. Opt. **47**, 291 (2005).
17. C. Rotschild, O. Cohen, O. Manela, M. Segev, and T. Carmon, "Solitons in nonlinear media with an infinite range of nonlocality: first observation of coherent elliptic solitons and of vortex-ring solitons," Phys. Rev. Lett. **95**, 213904 (2005).
18. C. Rotschild, B. Alfassi, O. Cohen, and M. Segev, "Long-range interactions between optical solitons," Nat. Phys. **2**, 769 (2006).

19. B. Alfassi, C. Rotschild, O. Manela, M. Segev, and D. N. Christodoulides, "Boundary force effects exerted on solitons in highly nonlocal nonlinear media," Opt. Lett. **32**, 154 (2007).
20. A. Minovich, D. Neshev, A. Dreischuh, W. Krolikowski, and Y. Kivshar, "Experimental reconstruction of nonlocal response of thermal nonlinear optical media," Opt. Lett., Doc. ID 78152 (in press).
21. Z. Xu, Y. V. Kartashov, and L. Torner, "Upper threshold for stability of multipole-mode solitons in nonlocal nonlinear media," Opt. Lett. **30**, 3171 (2005).

In many optical materials, a high-power light beam modifies the refractive index with the local light intensity. Nevertheless, in some materials the response can be highly nonlocal, so that the nonlinear contribution to the refractive index depends on the intensity distribution in the entire transverse plane, on a characteristic spatial scale that depends on the type of nonlocality. Different types of nonlocalities are encountered in actual materials, such as photorefractive, liquid crystals or plasmas. Such nonlocalities can have a profound effect on the properties of nonlinear excitations [1-4]. In two transverse dimensions, nonlocality stabilizes not only fundamental solitons, but also more complex structures, such as multipole-mode solitons, studied theoretically in Refs. [5-7] and observed in Ref. [8], rotating azimuthally modulated beams connecting multipole and vortex soliton families [6,9], and vortex-ring solitons [10-15]. Notice that, because of their particular shape, the latter tend to be highly prone to azimuthal instabilities in most local materials [16].

Vortex solitons with single topological charge have been recently observed in a self-focusing thermal medium [17]. In such materials long-range soliton interactions take place [18], while the geometry of the sample affects the beam trajectory [19] and determines the shape of the induced refractive index profile [20]. Thermal nonlinearities occur in many optical media, even though they become dominant only under specific conditions in materials with high enough light absorption coefficient, large thermo-optic effects, etc. Thermal nonlinearities are intrinsically nonlocal, with the range of nonlocality being determined by the geometry of the sample and its dimensions.

The fact that nonlocality stabilizes vortex solitons is well established [10-12]. However, the shape of the response function of the medium, which depends on the physical mechanism of nonlocality, and its inner scale, appears to be crucial factor to set a *maximal possible charge* of stable vortex solitons. For example, no limits have been found for the charge of stable nonlinear vortices in materials with a Gaussian-shaped response function [11], while in media described by a Helmholtz-type response function (such as liquid crystals) only single-charge vortex solitons are stable [12]. In this paper we discover that in thermal nonlinear medium with a cylindrical symmetry only vortex solitons with topological charge $m \leq 2$ can be stable under propagation. This finding is the outcome of a rigorous linear stability analysis and it is confirmed by accurate, direct simulations of the propagation of perturbed vortex soliton stationary solutions.

We consider the propagation of a laser beam along the $\xi$ axis in a focusing thermal medium of circular cross-section described by the following system of equations:

$$i\frac{\partial q}{\partial \xi} = -\frac{1}{2}\left(\frac{\partial^2 q}{\partial r^2} + \frac{1}{r}\frac{\partial q}{\partial r} + \frac{1}{r^2}\frac{\partial^2 q}{\partial \phi^2}\right) - qn,$$
$$\frac{\partial^2 n}{\partial r^2} + \frac{1}{r}\frac{\partial n}{\partial r} + \frac{1}{r^2}\frac{\partial^2 n}{\partial \phi^2} = -|q|^2. \tag{1}$$

Here $q = (k_0^2 r_0^4 \alpha |\beta| / \kappa n_0)^{1/2} A$ is the dimensionless light field amplitude; $n = k_0^2 r_0^2 \delta n / n_0$ is proportional to the nonlinear change $\delta n$ of the refractive index $n_0$; $\alpha, \beta, \kappa$ are the optical absorption, thermo-optic, and thermal conductivity coefficients, respectively; the radial and longitudinal coordinates $r, \xi$ are scaled to the characteristic beam radius $r_0$ and the diffraction length $k_0r_0^2$, respectively; $\phi$ is the azimuthal angle. We solved the system (1) with the boundary conditions $q,n|_{r\to R}=0$, where $R$ is the radius of the sample. Such boundary conditions are consistent with typical experimental conditions where a laser beam is focused to a typical radius of tens of microns and the rod radius amounts to a few millimeters. We assume that the boundary of the medium is kept at a fixed temperature. Notice that a similar physical model was considered by Kruglov and co-workers in Ref. [10], where the very fact that azimuthal instabilities of nonlinear vortex beams can be suppressed in thermal nonlinear media was predicted.

Under such conditions light launched along the axis of cylindrical rod and experiencing slight absorption, raises the temperature of the medium, while heat diffusion causes temperature redistribution in the entire sample that reaches a steady state after a relatively long time interval, which is proportional to $R^2$ and inversely proportional to the thermal conductivity coefficient. Increasing the temperature in a material with positive thermo-optic coefficient $\beta>0$ leads to a refractive index growth and facilitates light trapping in the heated regions, not only for simplest bell-shaped beams but also for topologically nontrivial beams carrying phase singularities.

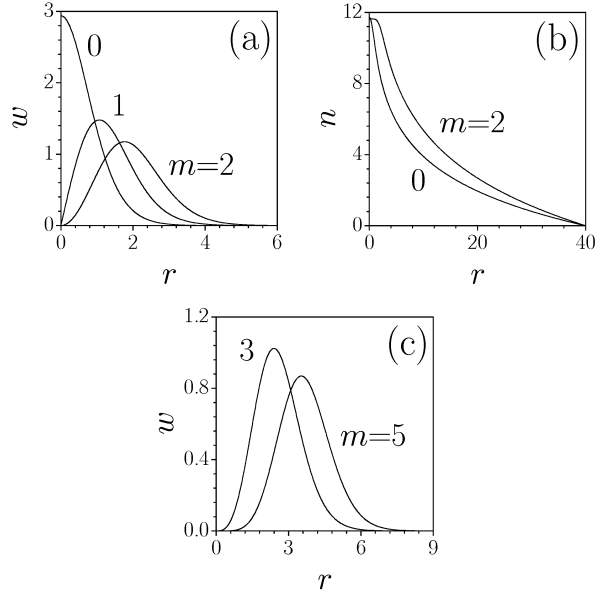

Fig. 1. (a) Profiles of solitons with topological charges $m=0$, $1$, and $2$ at $b=10$. (b) Refractive index profile for solitons with $m=0$ and $2$ at $b=10$. (c) Profiles of vortex solitons with $m=3$ and $5$ at $b=10$.

For radially symmetric excitations one can write an integral representation of solution of the thermo-conductivity equation as

$$n(r,\xi)=-\int_0^R G_0(r,\rho)|q(\rho,\xi)|^2\,d\rho,$$

where the response function of the thermal medium is given by $G_0(r,\rho)=\rho\ln(\rho/R)$ for $r<\rho$ and $G_0(r,\rho)=\rho\ln(r/R)$ for $r\geq\rho$. As mentioned above, the shape of the response function is set by the geometry of the sample and depends on its transverse extent. Note that this is in a sharp contrast to materials with Helmholtz [7,12] and Gaussian [11] response functions possessing an intrinsic nonlocality scale. In our computer simulations reported here we set $R = 40$, a value which closely resembles actual experimental conditions. However, it is worth stressing that all the results obtained remain qualitatively valid for other $R$ values.

We searched for stationary solutions of Eq. (1) in the form $q = w(r)\exp(ib\xi + im\phi)$, where $w(r)$ is a real function describing the profile of radially symmetric nonlinear wave solutions, $b$ is the propagation constant, and $m$ is the winding number, or topological charge. The profiles of fundamental solitons ($m = 0$) and vortices ($m \geq 1$) are found numerically with a standard relaxation algorithm. To analyze the soliton stability, we look for perturbed solutions in the form $q=[w(r)+u(r,\xi)\exp(ik\phi)+v^*(r,\xi)\exp(-ik\phi)]\exp(ib\xi+im\phi)$, where $u, v$ are the perturbation components that can grow with a complex rate $\delta$ upon propagation, and $k$ is the azimuthal perturbation index. Substitution of the light field in such form into Eq. (1) and linearization yields the eigenvalue problem:

$$i\delta u = -\frac{1}{2}\left(\frac{d^2 u}{dr^2} + \frac{1}{r}\frac{du}{dr} - \frac{(m+k)^2}{r^2}u\right) - w\Delta n_k - un + bu,$$
$$i\delta v = \frac{1}{2}\left(\frac{d^2 v}{dr^2} + \frac{1}{r}\frac{dv}{dr} - \frac{(m-k)^2}{r^2}v\right) + w\Delta n_k + vn - bv, \tag{2}$$

where

$$\Delta n_k = -\int_0^R G_k(r,\rho)w(\rho)[u(\rho) + v(\rho)]d\rho$$

is the refractive index perturbation corresponding to the azimuthal index $k$, while $G_k(r,\rho) = -(\rho/2k)[(r/\rho)^k - (r\rho/R^2)^k]$ at $r<\rho$ and $G_k(r,\rho)=-(\rho/2k)[(\rho/r)^k-(r\rho/R^2)^k]$ at $r \geq \rho$. We solved the eigenvalue problem (2) numerically.

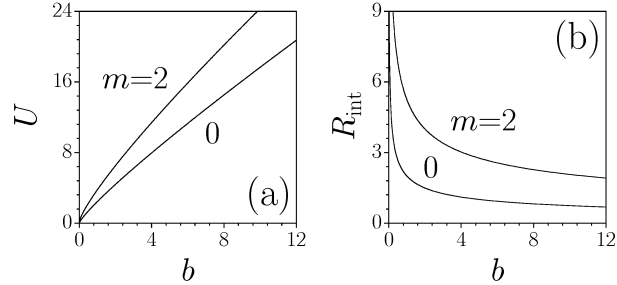

Fig. 2. Energy flow (a) and integral radius (b) versus propagation constant for solitons with different topological charges.

Representative profiles of solitons with different topological charges $m$ are shown in Fig. 1. Besides fundamental solitons, thermal media support ring-shaped vortex solitons of any charge for any propagation constant $b > 0$. At fixed $b$, the vortex-ring radius gradually increases with growth of $m$, while the peak amplitude decreases. The energy flow and the integral radius, defined as

$$U = 2\pi\int_0^\infty rw^2 dr, \qquad R_{\text{int}} = 2\pi U^{-1}\int_0^\infty r^2 w^2 dr,$$

have larger values for solitons with higher topological charges $m$ at fixed $b$. While soliton profiles are strongly localized (provided that $b$ is not too small), the temperature and the refractive index distributions are much broader and cover the whole cross-section of the rod. Such refractive index distribution forms a waveguide that traps light near the axis of the cylindrical sample. Notice that the refractive index profile features an almost flat plateau around the point $r = 0$ for vortex solitons with $m > 0$ (Fig. 1(b)), while for fundamental soliton it exhibits a cone-like profile instead. The width of the plateau increases with growth of the topological charge. Increasing $b$ results in a monotonic growth of the energy flow $U$ for any $m$ (Fig. 2(a)) and leads to a progressive soliton localization, so that the integral soliton radius monotonically decreases with $b$ (Fig. 2(b)).

Besides the simplest solitons shown in Fig. 1, we found for any value of topological charge $m$ a number of higher-order soliton families for which the field $w(r)$ has one or multiple nodes along the radial direction (not shown here). Our stability analysis has proved that all such solutions are *unstable*. This is in clear contrast, e.g., to media with a Gaussian response function where ring-like solitons with nodes can be stable [11].

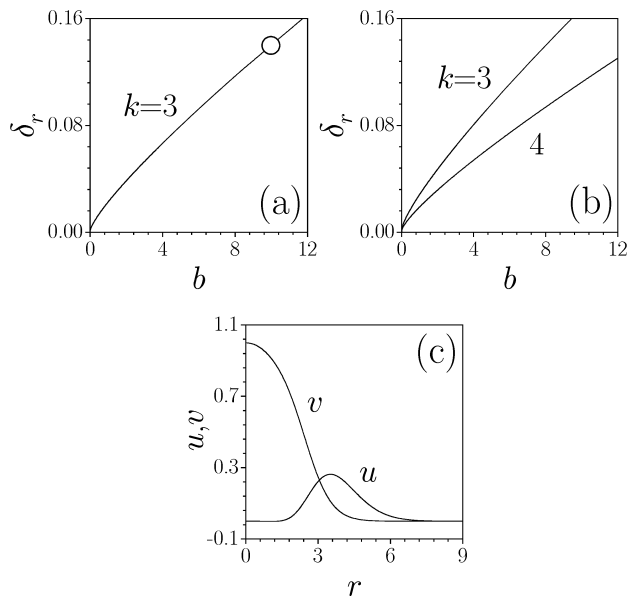

Fig. 3. Real part of the perturbation growth rate versus the propagation constant for vortex solitons with $m = 3$ (a) and $m = 4$ (b) for different azimuthal indices $k$. The point labeled by the circle in (a) corresponds to the perturbation profile shown in (c).

The rigorous linear stability analysis that we have conducted shows that fundamental solitons are stable in the entire existence domain, which is consistent with the so-called Vakhitov-Kolokolov stability criterion, since $dU/db > 0$ for any $b$ value. *The central result of this paper is that vortex solitons can be stable only if their topological charges fulfil $m \leq 2$.* Our results indicate that such vortex solitons are stable in the entire existence domain, irrespectively of the radius $R$ of the medium. Therefore, this is a rather unique constraint. For example, recall that in media with a Helmholtz-type response function only charge-one vortex solitons can be stable [12], while in the case of Gaussian response there are no restrictions on the charge of stable vortex solitons [11]. On physical grounds, the shrinking of the stability band with increasing topological charge in thermal media is related to the flattening of the refractive index profile, which causes the corresponding reduction of the focusing strength of the induced thermal lens. Notice also that, intuitively, the constraint mentioned above is somewhat similar to the restriction on the maximal number of solitons that can be packed into a stable train in one-dimensional nonlocal media with exponential response [21].

Stabilization of vortex solitons is possible because in thermal media a local increase of the beam intensity in one spatial point causes a refractive index modification in the entire sample cross-section, so that the induced refractive index does not feature a local maximum in that spatial point. All vortices with charges $m>2$ are unstable (see Fig. 1(c)) for examples of profiles of such vortices), and perturbations with rather specific azimuthal indices can quickly lead to their breakup. This is in contrast to other materials, where perturbations with multiple azimuthal indices might be almost equally destructive. That is, $m=3$ vortex solitons are unstable with respect to perturbations corresponding to $k=3$ (see Fig. 3(a) for dependence of the real part of the perturbation growth rate on the propagation constant $b$), while vortex solitons with topological charges $m=4,5$ are unstable with respect to perturbations with $k=3,4$ (Fig. 3(b)), etc. All such instabilities were found to be oscillatory, with $\delta_r, \delta_i \neq 0$, and in all cases the real part of the increment $\delta_r$ was found to grow with $b$.

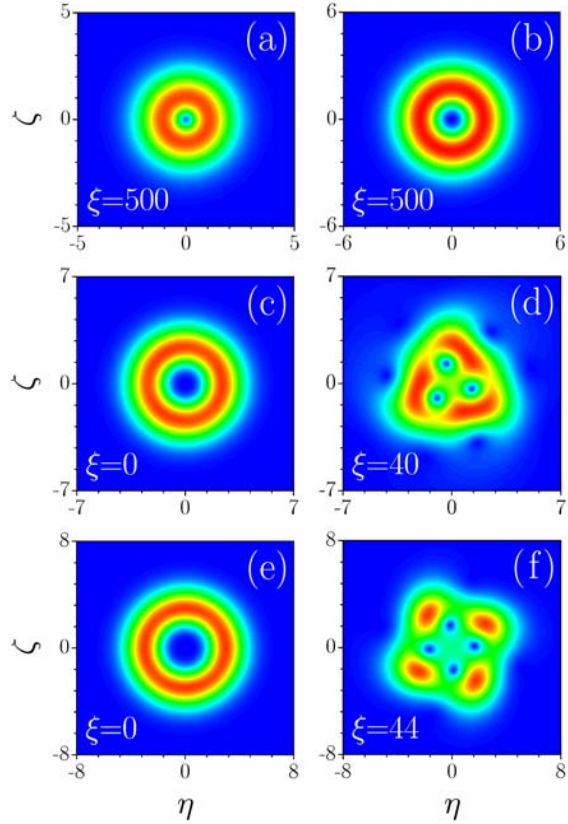

Fig. 4. Evolution of perturbed vortex solitons. Field modulus distributions are shown at different propagation distances for (a) $m=1$, (b) $m=2$, (c) and (d) $m=3$, (e) and (f) $m=4$.

Direct propagation of perturbed vortex solitons confirmed the above described predictions of the linear stability analysis. Vortices with $m = 1$ and $2$ at any $b$ value survive over indefinitely long distances even in the presence of strong broadband input noise added to the stationary solutions (see Figs. 4(a) and 4(b) showing perturbed $m = 1, 2$ vortices after their propagation over considerable distance). Vortices with $m = 3, 4$ decay in a specific fashion (see Figs. 4(c),(d) and 4(e),(f) where the initial multiplicative perturbations $[1 + 0.01\cos(3\phi)]$ and $[1 + 0.01\cos(4\phi)]$ were added into vortices with charge $3$ and $4$, respectively). In all cases the on-axis topological wavefront singularity self-unfolds into single-charge dislocations and one can clearly see the appearance of the individual singularities in the beam shape that rotate with distance and gradually move toward the periphery of the beam. The beams eventually transforms into fundamental solitons. This decay scenario is entirely consistent with the shape of the critical perturbation shown in Fig. 3(c) for $m = 3$ and $k = 3$. One can see that a growing $v$ component should result in increase of the optical field on the axis of the sample, while the $u$ component results in appearance of an azimuthal modulation.

Therefore, summarizing, we predicted that vortex solitons in thermal nonlinear media with a circular cross-section can be stable only if their winding number, or topological charge, verifies $m \leq 2$. This constraint on the maximal possible topological vortex charge is unique and is different from similar constraints encountered in other nonlocal materials. Importantly, note that we found a similar result in geometries with square cross-section (i.e., all vortices with $m > 2$ were found to be unstable in square samples, while vortices with $m \leq 2$ were stable). This suggests that the finding reported in here might be general for vortex-type solitons in materials with thermal nonlinearities.

## Acknowledgements

This work has been supported in part by the Government of Spain through the Ramon-y-Cajal program and through the grant TEC2005-07815.